\documentclass[aps, prl, reprint, preprintnumbers, superscriptaddress, longbibliography]{revtex4-2}
\usepackage{latexsym,epsfig,amssymb,amsfonts,amsmath,graphicx,bbm,braket,esdiff,dsfont,array,hyperref}
\pdfoutput=1
\hypersetup{
pdftitle={Comment on ``Relaxation Theory for Perturbed Many-Body Quantum Systems versus Numerics and Experiment''}, pdfauthor={Paul Secular}, pdfkeywords={computational, strongly-correlated, quantum, simulator, numerics, physics, tensor network, Bose-Hubbard, Hamiltonian, time evolution, dynamics, relaxation, simulation, TEBD, DMRG, MPS}, colorlinks=true, allcolors=blue}

\usepackage{xcolor}
\usepackage[percent]{overpic}
\usepackage[shortlabels]{enumitem}
\usepackage[english]{babel}
\usepackage[subrefformat=parens,labelformat=parens,caption=false]{subfig}

\usepackage{afterpage}

\addto\captionsenglish{
  \renewcommand{\figurename}{FIG.}
  \renewcommand{\tablename}{TABLE}
  
}

\usepackage{soul}

\makeatletter
\newcommand{\colorcaption}[2][]{
  \begingroup
  \renewcommand{\@caption@fignum@sep}{ (color online). }
  \caption[#1]{#2}
  \endgroup
}
\makeatother

\allowdisplaybreaks

\begin{document}

\title{Comment on ``Relaxation Theory for Perturbed Many-Body Quantum Systems versus Numerics and Experiment''}

\author{Paul Secular}
\email{paul@secular.me.uk}
\homepage{http://secular.me.uk/}
\affiliation{Department of Physics, University of Bath, Claverton Down, Bath BA2 7AY, United Kingdom}

\date{\today}

\maketitle

In Fig. 2 of a recent Letter \cite{dabelow_relaxation_2020}, Dabelow and Reimann compare their relaxation theory to data from Trotzky \emph{et al.}'s quantum simulator experiment \cite{trotzky_probing_2012}, finding an unexpected discrepancy. I explain this by pointing out that the quasi-local observable measured in the experiment is affected by the presence of a harmonic trapping potential \cite{trotzky_probing_2012} that is unaccounted for in the analytic calculation. I support this claim with quasi-exact numerics, and show that the theory gives accurate results if compared to a more appropriate local observable.

In their Examples section \cite{dabelow_relaxation_2020}, Dabelow and Reimann consider the Bose-Hubbard model (Eq. 6 of the Letter) for an infinite chain with onsite interaction energy $U$ and hopping amplitude $J$. On the other hand, Trotsky \emph{et al.} simulate a finite Bose-Hubbard chain in the presence of an external harmonic trap of strength $K$ \cite{trotzky_probing_2012}. In Fig. 2 of the Letter \cite{dabelow_relaxation_2020}, the authors consider the observable $A = \hat{n}_1$. Translational invariance means this can be taken as the density of any initially unoccupied (``odd'') site. The experimental data from Trotzky \emph{et al.}, however, are the ensemble average of the ``odd-site population'' $n_{\text{odd}}$ for a number of chains with different total particle number $N$ \cite{trotzky_probing_2012}. No expression for $n_{\text{odd}}$ is given in Ref. \cite{trotzky_probing_2012}, but it is reasonable to assume \footnote{Ref. \cite{schmitteckert_relaxation_2012} notes that Trotzky \emph{et al.} did not respond to a request for clarification on this point.} $n_{\text{odd}} = \sum_{j \in \text{odd}}{\braket{\hat{n}_j}} / N$, as in Refs. \cite{schmitteckert_relaxation_2012, urbanek_parallel_2016}. While $n_{\text{odd}} = \braket{\hat{n}_1}$ in a translationally invariant setting \cite{flesch_probing_2008}, this does not hold for Ref. \cite{trotzky_probing_2012}. Boundary \cite{flesch_probing_2008} and ensemble averaging \cite{trotzky_probing_2012} effects are small for the times considered here, but that of a trap can be significant \cite{flesch_probing_2008}.

\begin{figure}
	\includegraphics[width=8.6cm,keepaspectratio]{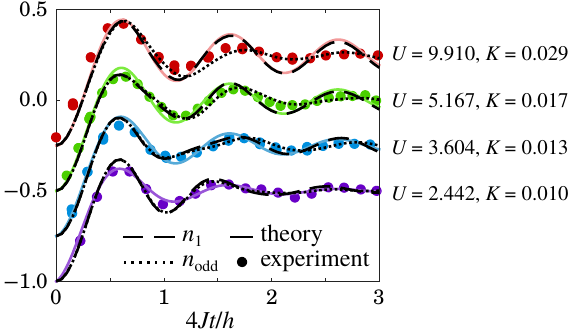}
	\colorcaption{\label{figPlot}$n_1$ and $n_{\text{odd}}$ for a 121-site chain with 43 particles computed using the parameters from Ref. \cite{trotzky_probing_2012}, compared with the theoretical and experimental \cite{trotzky_probing_nodate} values given in Fig. 2 of the Letter \cite{dabelow_relaxation_2020} (with the same vertical shifting).}
\end{figure}

Figure \ref{figPlot} shows numerical results for $n_{\text{odd}}$ and $\braket{\hat{n}_1}$ computed using a novel variant \footnote{This new variant of parallel TEBD will be described by the author in a forthcoming work.} of the time-evolving block decimation (TEBD) algorithm \cite{vidal_efficient_2003, vidal_efficient_2004}. To minimize the effects of the trap, $n_1 = \braket{\hat{n}_1}$ is taken to be the density of the most central odd site. The computed results for $n_1$ with $K \neq 0$ differ from those for $n_1$ and $n_{\text{odd}}$ with $K = 0$ (not shown) by $\lesssim 3\%$ for the times shown. However, it is clear that $n_{\text{odd}}$ is far more sensitive to the presence of the trap. My plots of $n_1$ should thus be a more appropriate benchmark for the relaxation theory. Indeed, the numerics for $n_1$ show excellent agreement with the theory for $U=9.9910$, but deviate for smaller $U$. This is consistent with the trend seen in Fig. 1 of the Letter \cite{dabelow_relaxation_2020}, in contrast to its Fig. 2. The close match between theory and experiment for small $U$ and $K$ is also worth noting. Trotzky \emph{et al.} point out that their tight-binding assumption breaks down for small $U$ \cite{trotzky_probing_2012}, which suggests the relaxation theory may better describe the Bose-Hubbard model with next-nearest neighbor hopping in this regime.

Incidentally, my results for $n_{\text{odd}}$ disagree with the numerics presented by Trotzky \emph{et al.} in their Supplementary Fig. 2 \cite{trotzky_probing_2012}. These can, however, be reproduced using values of $K$ that differ from Refs. \cite{trotzky_probing_nodate, trotzky_probing_2012}. This may partially explain the discrepancy found in Fig 2d of Ref. \cite{trotzky_probing_2012}. To independently validate my results, I carried out the same simulations using \hbox{TEBDOL} \cite{urbanek_parallel_2016}.

\begin{acknowledgments}
P.S. is grateful to Stephen R. Clark, Lennart Dabelow, and Sergey Dolgov for helpful discussions, and Miroslav Urbanek for help setting up TEBDOL, and for providing a bug fix. This research made use of the Balena High Performance Computing Service at the University of Bath, and built on Tensor Network Theory Library code written by Sarah Al-Assam and Chris Goodyer \cite {goodyer_tnt_nodate,
	al-assam_tensor_2017}. P.S. is funded by ClusterVision and the University of Bath.
\end{acknowledgments}


\begin{thebibliography}{13}%
	\makeatletter
	\providecommand \@ifxundefined [1]{%
		\@ifx{#1\undefined}
	}%
	\providecommand \@ifnum [1]{%
		\ifnum #1\expandafter \@firstoftwo
		\else \expandafter \@secondoftwo
		\fi
	}%
	\providecommand \@ifx [1]{%
		\ifx #1\expandafter \@firstoftwo
		\else \expandafter \@secondoftwo
		\fi
	}%
	\providecommand \natexlab [1]{#1}%
	\providecommand \enquote  [1]{``#1''}%
	\providecommand \bibnamefont  [1]{#1}%
	\providecommand \bibfnamefont [1]{#1}%
	\providecommand \citenamefont [1]{#1}%
	\providecommand \href@noop [0]{\@secondoftwo}%
	\providecommand \href [0]{\begingroup \@sanitize@url \@href}%
	\providecommand \@href[1]{\@@startlink{#1}\@@href}%
	\providecommand \@@href[1]{\endgroup#1\@@endlink}%
	\providecommand \@sanitize@url [0]{\catcode `\\12\catcode `\$12\catcode
		`\&12\catcode `\#12\catcode `\^12\catcode `\_12\catcode `\%12\relax}%
	\providecommand \@@startlink[1]{}%
	\providecommand \@@endlink[0]{}%
	\providecommand \url  [0]{\begingroup\@sanitize@url \@url }%
	\providecommand \@url [1]{\endgroup\@href {#1}{\urlprefix }}%
	\providecommand \urlprefix  [0]{URL }%
	\providecommand \Eprint [0]{\href }%
	\providecommand \doibase [0]{https://doi.org/}%
	\providecommand \selectlanguage [0]{\@gobble}%
	\providecommand \bibinfo  [0]{\@secondoftwo}%
	\providecommand \bibfield  [0]{\@secondoftwo}%
	\providecommand \translation [1]{[#1]}%
	\providecommand \BibitemOpen [0]{}%
	\providecommand \bibitemStop [0]{}%
	\providecommand \bibitemNoStop [0]{.\EOS\space}%
	\providecommand \EOS [0]{\spacefactor3000\relax}%
	\providecommand \BibitemShut  [1]{\csname bibitem#1\endcsname}%
	\let\auto@bib@innerbib\@empty
	\bibitem [{\citenamefont {Dabelow}\ and\ \citenamefont
		{Reimann}(2020)}]{dabelow_relaxation_2020}%
	\BibitemOpen
	\bibfield  {author} {\bibinfo {author} {\bibfnamefont {L.}~\bibnamefont
			{Dabelow}}\ and\ \bibinfo {author} {\bibfnamefont {P.}~\bibnamefont
			{Reimann}},\ }\bibfield  {title} {\bibinfo {title} {Relaxation {Theory} for
			{Perturbed} {Many}-{Body} {Quantum} {Systems} versus {Numerics} and
			{Experiment}},\ }\href {https://doi.org/10.1103/PhysRevLett.124.120602}
	{\bibfield  {journal} {\bibinfo  {journal} {Phys. Rev. Lett.}\ }\textbf
		{\bibinfo {volume} {124}},\ \bibinfo {pages} {120602} (\bibinfo {year}
		{2020})},\ \bibinfo {note} {publisher: American Physical Society}\BibitemShut
	{NoStop}%
	\bibitem [{\citenamefont {Trotzky}\ \emph {et~al.}(2012)\citenamefont
		{Trotzky}, \citenamefont {Chen}, \citenamefont {Flesch}, \citenamefont
		{McCulloch}, \citenamefont {Schollw{\"o}ck}, \citenamefont {Eisert},\ and\
		\citenamefont {Bloch}}]{trotzky_probing_2012}%
	\BibitemOpen
	\bibfield  {author} {\bibinfo {author} {\bibfnamefont {S.}~\bibnamefont
			{Trotzky}}, \bibinfo {author} {\bibfnamefont {Y.-A.}\ \bibnamefont {Chen}},
		\bibinfo {author} {\bibfnamefont {A.}~\bibnamefont {Flesch}}, \bibinfo
		{author} {\bibfnamefont {I.~P.}\ \bibnamefont {McCulloch}}, \bibinfo {author}
		{\bibfnamefont {U.}~\bibnamefont {Schollw{\"o}ck}}, \bibinfo {author}
		{\bibfnamefont {J.}~\bibnamefont {Eisert}},\ and\ \bibinfo {author}
		{\bibfnamefont {I.}~\bibnamefont {Bloch}},\ }\bibfield  {title} {\bibinfo
		{title} {Probing the relaxation towards equilibrium in an isolated strongly
			correlated one-dimensional {Bose} gas},\ }\href
	{https://doi.org/10.1038/nphys2232} {\bibfield  {journal} {\bibinfo
			{journal} {Nature Phys}\ }\textbf {\bibinfo {volume} {8}},\ \bibinfo {pages}
		{325} (\bibinfo {year} {2012})},\ \bibinfo {note} {number: 4 Publisher:
		Nature Publishing Group}\BibitemShut {NoStop}%
	\bibitem [{Note1()}]{Note1}%
	\BibitemOpen
	\bibinfo {note} {Ref. \cite {schmitteckert_relaxation_2012} notes that
		Trotzky \protect \emph {et al.} did not respond to a request for
		clarification on this point.}\BibitemShut {Stop}%
	\bibitem [{\citenamefont
		{Schmitteckert}(2012)}]{schmitteckert_relaxation_2012}%
	\BibitemOpen
	\bibfield  {author} {\bibinfo {author} {\bibfnamefont {P.}~\bibnamefont
			{Schmitteckert}},\ }\bibfield  {title} {\bibinfo {title} {On the relaxation
			toward equilibrium in an isolated strongly correlated one-dimensional {Bose}
			gas},\ }\href {https://doi.org/10.1088/0031-8949/2012/T151/014059} {\bibfield
		{journal} {\bibinfo  {journal} {Phys. Scr.}\ }\textbf {\bibinfo {volume}
			{T151}},\ \bibinfo {pages} {014059} (\bibinfo {year} {2012})},\ \bibinfo
	{note} {publisher: IOP Publishing}\BibitemShut {NoStop}%
	\bibitem [{\citenamefont {Urbanek}\ and\ \citenamefont
		{Sold{\'a}n}(2016)}]{urbanek_parallel_2016}%
	\BibitemOpen
	\bibfield  {author} {\bibinfo {author} {\bibfnamefont {M.}~\bibnamefont
			{Urbanek}}\ and\ \bibinfo {author} {\bibfnamefont {P.}~\bibnamefont
			{Sold{\'a}n}},\ }\bibfield  {title} {\bibinfo {title} {Parallel
			implementation of the time-evolving block decimation algorithm for the
			{Bose}{\textendash}{Hubbard} model},\ }\href
	{https://doi.org/https://doi.org/10.1016/j.cpc.2015.10.016} {\bibfield
		{journal} {\bibinfo  {journal} {Comput. Phys. Commun.}\ }\textbf {\bibinfo
			{volume} {199}},\ \bibinfo {pages} {170 } (\bibinfo {year}
		{2016})}\BibitemShut {NoStop}%
	\bibitem [{\citenamefont {Flesch}\ \emph {et~al.}(2008)\citenamefont {Flesch},
		\citenamefont {Cramer}, \citenamefont {McCulloch}, \citenamefont
		{Schollw{\"o}ck},\ and\ \citenamefont {Eisert}}]{flesch_probing_2008}%
	\BibitemOpen
	\bibfield  {author} {\bibinfo {author} {\bibfnamefont {A.}~\bibnamefont
			{Flesch}}, \bibinfo {author} {\bibfnamefont {M.}~\bibnamefont {Cramer}},
		\bibinfo {author} {\bibfnamefont {I.~P.}\ \bibnamefont {McCulloch}}, \bibinfo
		{author} {\bibfnamefont {U.}~\bibnamefont {Schollw{\"o}ck}},\ and\ \bibinfo
		{author} {\bibfnamefont {J.}~\bibnamefont {Eisert}},\ }\bibfield  {title}
	{\bibinfo {title} {Probing local relaxation of cold atoms in optical
			superlattices},\ }\href {https://doi.org/10.1103/PhysRevA.78.033608}
	{\bibfield  {journal} {\bibinfo  {journal} {Phys. Rev. A}\ }\textbf {\bibinfo
			{volume} {78}},\ \bibinfo {pages} {033608} (\bibinfo {year} {2008})},\
	\bibinfo {note} {publisher: American Physical Society}\BibitemShut {NoStop}%
	\bibitem [{\citenamefont {Trotzky}\ \emph {et~al.}()\citenamefont {Trotzky},
		\citenamefont {Chen}, \citenamefont {Flesch}, \citenamefont {McCulloch},
		\citenamefont {Schollw{\"o}ck}, \citenamefont {Eisert},\ and\ \citenamefont
		{Bloch}}]{trotzky_probing_nodate}%
	\BibitemOpen
	\bibfield  {author} {\bibinfo {author} {\bibfnamefont {S.}~\bibnamefont
			{Trotzky}}, \bibinfo {author} {\bibfnamefont {Y.-A.}\ \bibnamefont {Chen}},
		\bibinfo {author} {\bibfnamefont {A.}~\bibnamefont {Flesch}}, \bibinfo
		{author} {\bibfnamefont {I.~P.}\ \bibnamefont {McCulloch}}, \bibinfo {author}
		{\bibfnamefont {U.}~\bibnamefont {Schollw{\"o}ck}}, \bibinfo {author}
		{\bibfnamefont {J.}~\bibnamefont {Eisert}},\ and\ \bibinfo {author}
		{\bibfnamefont {I.}~\bibnamefont {Bloch}},\ }\href
	{http://arxiv.org/abs/1101.2659} {\bibinfo {title} {Probing the relaxation
			towards equilibrium in an isolated strongly correlated {1D} {Bose}
			gas}}\BibitemShut {NoStop}%
	\bibitem [{Note2()}]{Note2}%
	\BibitemOpen
	\bibinfo {note} {This new variant of parallel TEBD will be described by the author in a forthcoming work.}\BibitemShut {Stop}%
	\bibitem [{\citenamefont {Vidal}(2003)}]{vidal_efficient_2003}%
	\BibitemOpen
	\bibfield  {author} {\bibinfo {author} {\bibfnamefont {G.}~\bibnamefont
			{Vidal}},\ }\bibfield  {title} {\bibinfo {title} {Efficient {Classical}
			{Simulation} of {Slightly} {Entangled} {Quantum} {Computations}},\ }\href
	{https://doi.org/10.1103/PhysRevLett.91.147902} {\bibfield  {journal}
		{\bibinfo  {journal} {Phys. Rev. Lett.}\ }\textbf {\bibinfo {volume} {91}},\
		\bibinfo {pages} {147902} (\bibinfo {year} {2003})}\BibitemShut {NoStop}%
	\bibitem [{\citenamefont {Vidal}(2004)}]{vidal_efficient_2004}%
	\BibitemOpen
	\bibfield  {author} {\bibinfo {author} {\bibfnamefont {G.}~\bibnamefont
			{Vidal}},\ }\bibfield  {title} {\bibinfo {title} {Efficient {Simulation} of
			{One}-{Dimensional} {Quantum} {Many}-{Body} {Systems}},\ }\href
	{https://doi.org/10.1103/PhysRevLett.93.040502} {\bibfield  {journal}
		{\bibinfo  {journal} {Phys. Rev. Lett.}\ }\textbf {\bibinfo {volume} {93}},\
		\bibinfo {pages} {040502} (\bibinfo {year} {2004})}\BibitemShut {NoStop}%
	\bibitem [{\citenamefont
		{Schollw{\"o}ck}(2011)}]{schollwock_density-matrix_2011}%
	\BibitemOpen
	\bibfield  {author} {\bibinfo {author} {\bibfnamefont {U.}~\bibnamefont
			{Schollw{\"o}ck}},\ }\bibfield  {title} {\bibinfo {title} {The density-matrix
			renormalization group in the age of matrix product states},\ }\href
	{https://doi.org/10.1016/j.aop.2010.09.012} {\bibfield  {journal} {\bibinfo
			{journal} {Ann. Phys. (N. Y.)}\ }\bibinfo {series} {January 2011 {Special}
			{Issue}},\ \textbf {\bibinfo {volume} {326}},\ \bibinfo {pages} {96}
		(\bibinfo {year} {2011})}\BibitemShut {NoStop}%
	\bibitem [{\citenamefont {Goodyer}()}]{goodyer_tnt_nodate}%
	\BibitemOpen
	\bibfield  {author} {\bibinfo {author} {\bibfnamefont {C.}~\bibnamefont
			{Goodyer}},\ }\href
	{http://www.hector.ac.uk/cse/distributedcse/reports/UniTNT/UniTNT/index.html}
	{\bibinfo {title} {{TNT} {Library} : {Tensor} {Manipulation} and
			{Storage}}}\BibitemShut {NoStop}%
	\bibitem [{\citenamefont {Al-Assam}\ \emph {et~al.}(2017)\citenamefont
		{Al-Assam}, \citenamefont {Clark},\ and\ \citenamefont
		{Jaksch}}]{al-assam_tensor_2017}%
	\BibitemOpen
	\bibfield  {author} {\bibinfo {author} {\bibfnamefont {S.}~\bibnamefont
			{Al-Assam}}, \bibinfo {author} {\bibfnamefont {S.~R.}\ \bibnamefont
			{Clark}},\ and\ \bibinfo {author} {\bibfnamefont {D.}~\bibnamefont
			{Jaksch}},\ }\bibfield  {title} {\bibinfo {title} {The tensor network theory
			library},\ }\href {https://doi.org/10.1088/1742-5468/aa7df3} {\bibfield
		{journal} {\bibinfo  {journal} {J. Stat. Mech.}\ }\textbf {\bibinfo {volume}
			{2017}},\ \bibinfo {pages} {093102} (\bibinfo {year} {2017})}\BibitemShut
	{NoStop}%
\end{thebibliography}
\end{document}